# Low Energy Excitations in a [4]Triangulene Honeycomb Antiferro-magnet

Aidan Delgado,[†,‡] Carolin Dusold,[†,‡] James Wu,[‡,‡] Jingwei Jiang,[‡] Christina M. Dadich,[†] Adam Cronin,[†] Steven G. Louie,[‡,¶]* Felix R. Fischer,[†,¶,§,f]*

[†] Department of Chemistry, University of California, Berkeley, CA 94720, USA.

[‡] Department of Physics, University of California, Berkeley, CA 94720, USA.

[¶] Materials Sciences Division, Lawrence Berkeley National Laboratory, Berkeley, CA 94720, USA.

[§] Kavli Energy NanoScience Institute at the University of California Berkeley and the Lawrence Berkeley National Laboratory, Berkeley, California 94720, USA

[f] Bakar Institute of Digital Materials for the Planet, Division of Computing, Data Science, and Society, University of California Berkeley; Berkeley, CA 94720, USA.

**ABSTRACT:** Carbon-based synthetic lattices offer a versatile platform for the realization and control of correlated quantum effects at the nanoscale. Advances in on-surface synthesis have facilitated the fabrication of increasingly complex molecular frameworks, opening new avenues for the exploration and engineering of custom tailored electronic and magnetic phases. Here, we report the design, on-surface synthesis, and characterization of a two-dimensional (2D) covalent organic framework (COF) assembled from $D_{3h}$ symmetric $S = {}^{3}/_{2}$ [4]triangulene building blocks arranged in a honeycomb structure. Combined scanning probe microscopy and spectroscopy, corroborated by first-principles density functional theory (DFT) and GW calculations, establish an antiferromagnetic (AFM) insulating ground state. Low energy spectroscopy resolves spin excitations in both one-dimensional (1D) chains and extended 2D networks, providing direct insight into the emergent spin dynamics and establishing a modular platform to explore quantum magnetism in a π-conjugated organic system. 

## INTRODUCTION

Magnetic order in crystalline solids arises from correlated alignment of electron spins mediated by strong electron-electron interactions. Characterization of these ordered phases sheds light on the many-body nature of a material. Among the most fundamental collective excitations of a magnetically ordered system are spin excitations, the quantized modes of spin waves, whose spectrum directly encodes the exchange interactions, lattice structure, and anisotropy of the underlying magnetic Hamiltonian. In antiferromagnetic (AFM) lattices, where neighboring spins have antiparallel alignment, the spin excitation spectrum reflects the interplay of competing exchange pathways, geometric frustration, and quantum fluctuations that can stabilize exotic ground states ranging from spin liquids to valence-bond solids.[1, 2] Probing these excitations provides a spectroscopic fingerprint of the magnetic Hamiltonian itself. While spin excitations have been detected in inorganic van der Waals (vdW) magnets,[3, 4] their observation in a purely organic 2D framework has remained elusive. COFs offer a bottom-up synthetic route to 2D quantum materials in which atomic connectivity, symmetry, and electronic structure are encoded at the molecular level.[5-8] Theoretical work predicting organic topological insulators and quantum anomalous Hall phases in 2D organometallic lattices,[9, 10] in addition to studies of intrinsic magnetism in COFs,[11] has established designer topological and magnetic quantum phases in MOF/COF architectures as an emerging field.[12] This intrinsic chemical modularity, absent in conventional inorganic 2D materials whose in-plane lattices remain largely immutable,[13-16] affords rational control over lattice structure, per-site unit spin magnitude, and exchange interactions, establishing COFs as a compelling platform for the design of organic quantum materials.

Guided by computational models[17-21] we identified [*n*]triangulenes (*n* is the number of annulated benzene rings along a zigzag edge),[22-27] a family of open-shell polycyclic aromatic hydrocarbons (PAHs), as competent building blocks for extended magnetic lattices. While experimental efforts have largely focused on 1D spin chains,[28-31] theory predicts that 2D lattices of high-spin triangulenes can host a wide range of phases, including second-order topological insulators,[32, 33] excitonic insulators,[17] and magnetically ordered states proximate to quantum spin liquid phases.[19, 34, 35] Here, we report the realization of a high-spin 2D-[4]triangulene covalent organic framework (2D-[4]TCOF) on Au(111). Low-temperature scanning tunneling microscopy (STM) and spectroscopy (STS), in conjunction with first-principles calculations, reveal an antiferromagnetically (AFM) ordered ground state that can be mapped onto a 2D Heisenberg model for the honeycomb lattice. The details of the electronic structure of 2D-[4]TCOF, along with the low-energy spin excitations measured by STS are supported by density functional theory (DFT)[36, 37] and GW[38, 39] simulations. These results establish extended π-conjugated COFs as a viable platform for high-spin quantum magnetism and highlight their potential for the bottom-up design of 2D organic quantum materials for spin-based technologies.

## RESULTS/DISCUSSION

**Design of a [4]TCOF honeycomb lattice.** Our strategy for engineering magnetic ordering in high-spin 2D [*n*]triangulene frameworks is based on a simple nearest neighbor (NN) tight binding (TB) model. The ground state spin quantum number (*S*) of all-carbon [*n*]triangulene units can be predicted by Ovchinnikov's rule[40] and Lieb's theorem[41] for bipartite lattices ($S = ½ (|N_A - N_B|)$ where $N_A$ and $N_B$ represent the number of atoms residing on the A and B graphene sublattice sites). DFT-GGA calculations[42] confirm the ground state for the $D_{3h}$ symmetric [4]triangulene ([4]T) is best represented as an open-shell quartet ($S = 3/2$) (Figures 1A,B).[24] The unpaired spins reside in a set of three singly occupied molecular orbitals (SOMOs) that extend across the π-system. A map of the total spin-density reveals a prominent localization at C-atoms lining the zigzag edges of the [4]T core, complemented by lesser contributions from atoms at the center and vertices. These three singly occupied near-degenerate zero-modes are reminiscent of the orbital coordination number of a site in a diatomic honeycomb lattice. Fused along its three vertices, [4]T gives rise to an orbital valence bond solid, the 2D-[4]TCOF, depicted in Figure 1C ($a_1$ and $a_2$ are the lattice vectors of the honeycomb lattice unit cell containing a pair of [4]T superatoms). Our DFT calculations predict an energetically favored AFM phase with three spin-polarized frontier bands, which are also spatially polarized on A/B [4]T units (Figure 1D). We therefore use an isotropic 2D Heisenberg model, where individual [4]T units are described as $S = 3/2$ sites interacting through antiferromagnetic exchange on a honeycomb lattice. This description is supported by the exact diagonalization of the Hubbard model for covalently linked [4]triangulene dimers, which shows that the low-energy many-body spectrum maps onto that of two AFM coupled $S = 3/2$ units.[18] In the absence of an external magnetic field the Hamiltonian takes the form of a bilinear-biquadratic-bicubic (BLBQBC) exchange term

$$\mathcal{H}_{\mathrm{BLBQBC}} = \Sigma_{\langle ij \rangle} J \left( \vec{S_i} \cdot \vec{S_j} + \beta_2 (\vec{S_i} \cdot \vec{S_j})^2 + \beta_3 (\vec{S_i} \cdot \vec{S_j})^3 \right) \quad (1)$$

where the sum over $\langle ij \rangle$ is over all bonds of sites *i* and *j* in the honeycomb lattice of [4]T superatoms, *J* the Heisenberg exchange coupling, $\vec{S_i}$ and $\vec{S_j}$ are the corresponding spin operators on sites *i* and *j*, and $\beta_2$, $\beta_3$ are parameters that respectively determine the strength of the biquadratic and bicubic terms relative to the bilinear term (Figure 1D).[18, 28] At the classical level, a phase diagram for the 2D honeycomb lattice BLBQBC model with $\beta_3 = 0$ (BLBQ) predicts four magnetic phases: a ferromagnetic (FM) phase, an antiferromagnetic phase (AFM), and the nematic ferro- (FQ) and antiferromagnetic quadrupolar (AFQ) phases (Figure 1E).

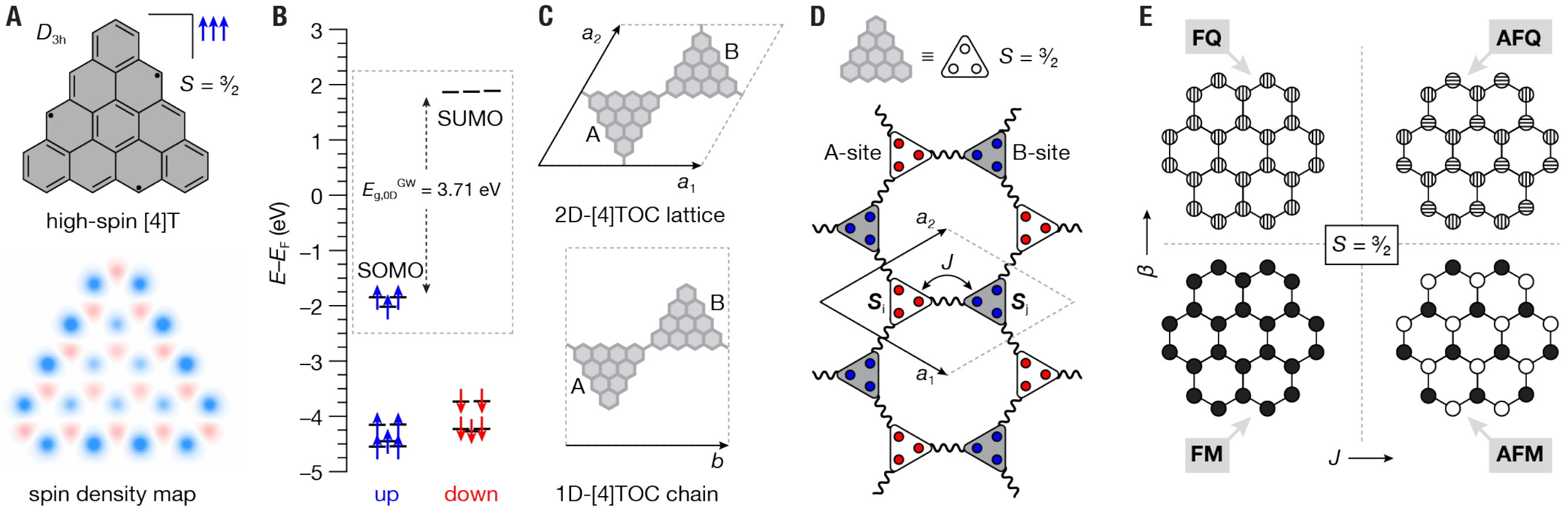

**Figure 1.** Assembly of high-spin [4]triangulenes into low dimensional lattices. (A) High-spin ($S = 3/2$) ground state of the $D_{3h}$ symmetric [4]triangulene (top). Spatial distribution of the net local spin density ($\rho(r)^{\uparrow} - \rho(r)^{\downarrow}$), integrated over all occupied states (bottom) with blue denoting the majority spin density. (B) DFT-GW energy level diagram showing the set of three SOMO (up-spin) and near-degenerate SUMO (down-spin) states of an isolated [4]triangulene core, with $E_F$ set at the middle of the SUMO-SOMO gap ($E_{g,0D}$). (C) Schematic representation of the crystallographic unit cell of a corner-linked honeycomb lattice (top) and a linear *poly*-[4]triangulene chain (bottom) formed from $S = 3/2$ high-spin [4]triangulene cores. (D) 2D Heisenberg model for corner-linked $S = 3/2$ [4]triangulene cores on a honeycomb lattice (empty and filled triangles denote A- and B-sites of honeycomb lattice formed by the cores, blue and red circles denote spin up and spin down electrons, sinusoid lines indicate hopping interactions that couple electrons on neighboring triangulenes). (E) Classical phase diagram for a 2D bilinear-biquadratic (BLBQ) spin model with interaction parameters *J* and *β* of corner-linked $S = 3/2$ [4]triangulene cores on a honeycomb lattice. The ferromagnetic (FM) and antiferromagnetic (AFM) phases are indicated by filled and open circles. The nematic ferro- and antiferromagnetic quadrupolar phases (FQ and AFQ) are highlighted by stripe patterns.

**Synthesis of molecular precursors and surface-assisted growth of [4]TCOF.** Guided by this idea we designed two competent molecular precursors for the reticular growth of a [4]triangulene COF ([4]TCOF), the benzo[*c*]naptho[2,1-*p*]chrysenes **1a** and **1b**. The synthesis is depicted in Figure 2A. *Suzuki-Miyaura* cross-coupling of benzene-1,3,5-triyltriboronic ester **2**[43] with an excess of 4-bromo-1-iodo(prop-1-en-1-yl)benzene **3** yields the tribromide **4** in 86% yield. Photocyclization of **4** in the presence of a stochiometric oxidant yields the tribrominated benzo[*c*]naphtho[2,1-*p*]chrysene core **1a** in 84%.[44] An aromatic *Finkelstein* reaction gives access to the corresponding triiodinated precursor for [4]TCOF **1b** in 86% yield.[45] Analytically pure samples of **1a** and **1b** suitable for ultrahigh- vacuum (UHV) deposition were isolated as racemic mixtures of slowly interconverting helical diastereoisomers (*PPP*/*MMM* and *PPM*/*MMP*).

[4]TCOFs were grown on Au(111)/mica films by sublimation of **1a**,**b** in UHV onto a clean Au(111) surface held at 297 K. Supporting Information Figure S1A,B shows constant-current STM topographic images of a sub-monolayer coverage of **1a** and **1b** on Au(111) at $T = 4.5$ K. Slow annealing of surfaces decorated with the tribrominated precursor **1a** to 473 K gave rise to highly irregular networks that hint at a competition between the desired *Ullmann*-type

coupling of aryl halides along the vertices and the surface catalyzed cyclodehydrogenation of methyl groups that establishes the extended zigzag edges in the [4]triangulene core (Supporting Information Figure S1C,D). To kinetically decouple these two reaction steps, the *Ullmann*-type C–C coupling and the thermal cyclodehydrogenation, we resorted to the more reactive triiodinated precursor **1b**.[46, 47] Molecule-decorated surfaces, > 0.9 monolayer (ML) **1b** on Au(111), were first annealed at 333 K for 12 h before slowly raising the surface temperature to 523 K (20 min) (Figure 2B). Figure 2C shows a representative constant-current STM image of a 2D covalent network of [4]triangulenes on Au(111) revealing a statistical distribution of 5-membered (18%), 6-membered (71%), and 7-membered rings 11%), respectively ($N$ = 3512, Supporting Information Figure S2).

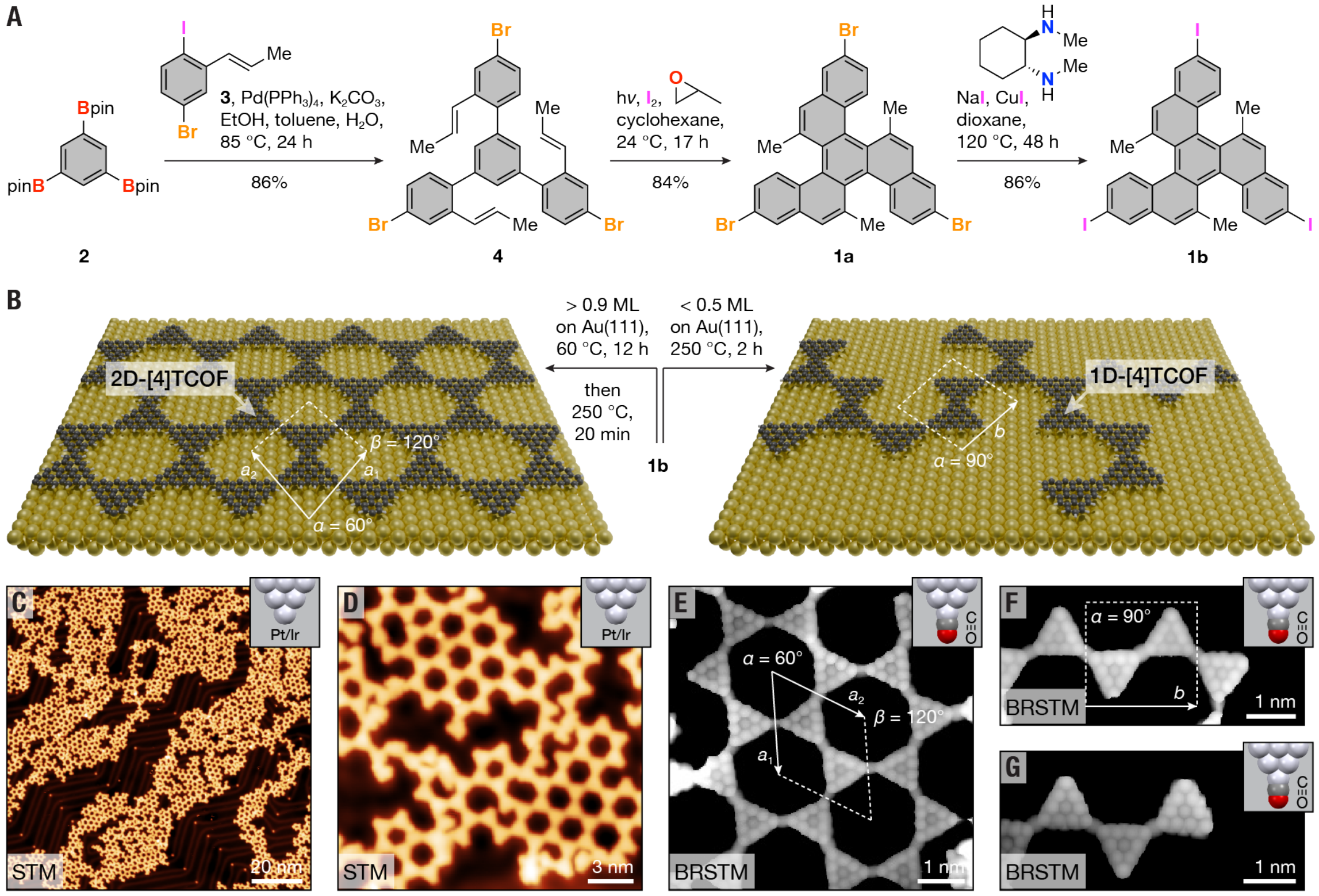

**Figure 2.** Bottom-up synthesis of low-dimensional [4]TCOFs. (A) Synthesis of molecular precursors **1a**,**b**. (B) Schematic representation of the on-surface growth of 2D-[4]TCOFs (left) and 1D-[4]TCOF chains (right) from molecular precursors **1a**,**b**. (C) STM topographic image of a 2D-[4]TCOF network on Au(111) ($V_s$ = –800 mV, $I_t$ = 10 pA). (D) STM topographic image of a of crystalline island of 2D-[4]TCOF formed by annealing a > 0.9 ML of **1b** on Au(111) first to $T_1$ =333 K and then to $T_2$ = 523 K ($V_s$ = 200 mV, $I_t$ = 20 pA). (E–G) Constant height BRSTM image of a 2D-[4]TCOF honeycomb lattice and corner-sharing 1D-[4]TCOF chains. [4]triangulene vertices not involved in C–C bonding are terminated by H-atoms ($V_s$ = 10 mV, $V_{ac}$ = 10 mV, $f$ =455 Hz, CO-functionalized tip). All STM experiments were performed at $T$ = 4.5 K.

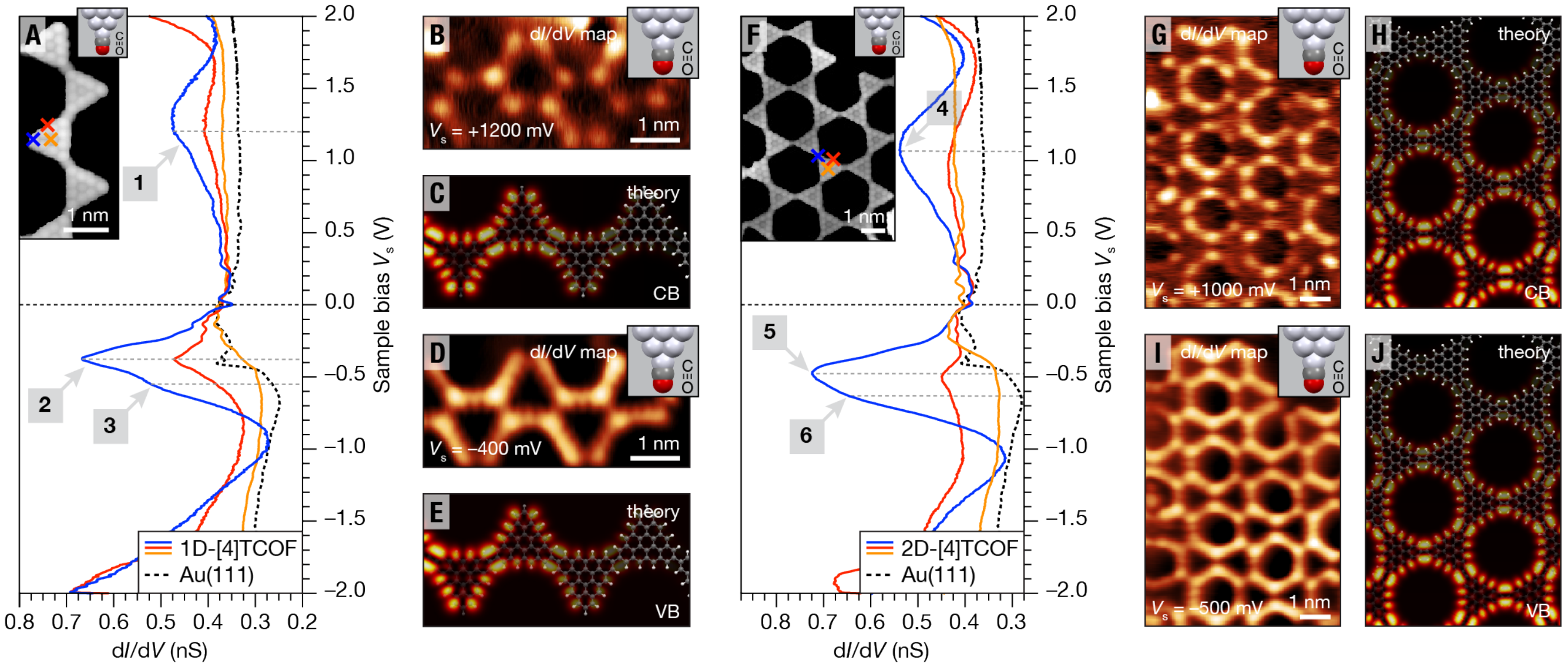

**Figure 3.** Electronic structure of 1D-[4]TCOF and 2D-[4]TCOF. (A) d$I$/d$V$ point spectroscopy of 1D-[4]TCOF chains on Au(111) recorded at the positions marked in the BRSTM inset (blue, red, and orange crosses; spectroscopy: $V_{ac}$ = 10 mV, $f$ = 455 Hz; constant-height imaging: $V_s$ = 10 mV, $V_{ac}$

= 10 mV, $f$ = 455 Hz, CO-functionalized tip). (B) Constant height d$I$/d$V$ map recorded at a sample voltage bias of $V_s$ = +1200 mV ($V_{ac}$ = 10 mV, $f$ = 455 Hz, CO-functionalized tip). (C) DFT-GGA calculated LDOS map for a freestanding of 1D-[4]TCOF at energies corresponding to the CB edge. (D) Constant height d$I$/d$V$ map recorded at a sample voltage bias of $V_s$ = –400 mV ($V_{ac}$ = 10 mV, $f$ = 455 Hz, CO-functionalized tip). (E) DFT-GGA calculated LDOS map for a freestanding of 1D-[4]TCOF at energies corresponding to the VB edge. (F) d$I$/d$V$ point spectroscopy of 2D-[4]TCOF on Au(111) recorded at the positions marked in the BRSTM inset (blue, red, and orange crosses; spectroscopy: $V_{ac}$ = 10 mV, $f$ = 455 Hz; constant-height imaging: $V_s$ = 10 mV, $V_{ac}$ = 10 mV, $f$ = 455 Hz, CO-functionalized tip). (G) Constant height d$I$/d$V$ map recorded at a sample voltage bias of $V_s$ = +1000 mV ($V_{ac}$ = 10 mV, $f$ = 455 Hz, CO-functionalized tip). (H) DFT-GGA calculated LDOS map for a freestanding of 2D-[4]TCOF at energies corresponding to the CB edge. (I) Constant height d$I$/d$V$ map recorded at a sample voltage bias of $V_s$ = –500 mV ($V_{ac}$ = 10 mV, $f$ = 455 Hz, CO-functionalized tip). (J) DFT-GGA calculated LDOS map for a freestanding of 2D-[4]TCOF at energies corresponding to the VB edge. All STM experiments were performed at $T$ = 4.5 K. Theoretical LDOS maps are sampled at a height of 4.0 Å above the atomic plane of the [4]TCOF.

Long range order of bottom-up synthesized [4]TCOFs resulting from **1b** remains low and crystallinity is confined to small, spatially isolated domains (Figure 2D). Bond-resolved STM (BRSTM) reveals that the 2D covalent network is formed by equilateral triangles featuring atomically smooth zigzag edges with apparent base length, and $z$-height of 0.85 nm ± 0.01 nm and 0.14 nm ± 0.02 nm, respectively, consistent with the formation of the fully cyclodehydrogenated [4]triangulene core (Figure 2E, Supporting Information Figure S3). Large area STM images reveal that the cross-linking in 2D-[4]TCOFs prepared from the triiodinated molecular precursor **1b** almost exclusively (> 99%) feature the desired C–C bonding between [4]triangulene vertices (Figure 2D) rather than unselective fusions involving the highly reactive zigzag edges commonly observed for **1a** (Supporting Information Figure S1C,D).

To access samples of the corresponding 1D-[4]TCOF chains, we gradually annealed Au(111) surfaces decorated with sub-monolayer coverages (< 0.5 ML) of **1b** to 523 K over the course of 2 h. The lower monomer density and faster annealing protocol favor the formation of an amorphous network of branched 1D-[4]TCOF chains (< 10 monomer units per chain segment). The threefold symmetry of the monomer gives rise to a statistical mixture of *cis*- and *trans*-linkages reflected in a sequence of linear and kinked segments along the polymer backbone (Supporting Information Figure S4). BRSTM images (Figure 2F) suggest that carbon centered radicals (resulting from the thermal activation of C–I bonds) lining the vertices of [4]triangulenes that failed to undergo *Ullmann*-type coupling as part of the polymerization step are terminated by H-atoms generated during the cyclodehydrogenation process.[48]

**Electronic structure characterization.** We characterized the local electronic structure of 1D-[4]TCOF chains and 2D-[4]TCOF honeycomb lattices using scanning tunneling spectroscopy (STS). Figure 3A shows representative d$I$/d$V$ spectra recorded on a 1D-[4]TCOF chain at the positions marked in the inset. Differential conductance spectra acquired at the center of [4]triangulene units were largely featureless and are dominated by the underlying Au(111) surface state. In contrast, spectra recorded at the junction between monomer units or along the zigzag edges showed three prominent features close to the Fermi level ($E_F$). A broad peak at positive bias centered at $V_s$ = +1.25 ± 0.05 V (*feature* 1) is mirrored by a narrow prominent peak at $V_s$ = –0.35 ± 0.05 V (*feature* 2) at negative bias that tails off into a shoulder at $V_s$ = –0.55 ± 0.05 V (*feature* 3). Bracketing the $E_F$ *features* 1 and 2 define a sizeable semiconducting gap estimated at $E_{g,1D}$ ~ 1.60 V (based on peak-to-peak separation). Differential conductance maps recorded across a wide bias window, +2.00 V > $V_s$ > –2.00 V, show prominent signatures of localized states that can be mapped onto the electronic structure of 1D-[4]TCOF chains (Supporting Information Figure S5). A distinctive nodal pattern, a dark featureless [4]triangulene core, contrasted by two bright lobes along the junction between monomer units, was observed in d$I$/d$V$ maps recorded at biases corresponding to *feature* 1 (Figure 3B). Residual signal intensity, albeit with diffuse contrast, follows the outline of zigzag edges. Differential conductance maps recorded at biases corresponding to *features* 2 and 3 show similar patterns, a low contrast at the [4]triangulene cores bordered by a series of bright lobes lining zigzag edges, and the covalent bonds forming the junction interface (Figure 3D). Experimental d$I$/d$V$ maps for an AFM ground state were reproduced by theory based on *ab-initio* DFT calculations under the generalized gradient approximation (GGA) for the exchange correlation functional. The local density of states (LDOS) corresponding to the conduction (CB) and valence band (VB) edges (Figures 3C,E) of a freestanding 1D-[4]TCOF chain are in good agreement with differential conductance maps recorded at biases corresponding to *features* 1 and 2, respectively.[49]

Having explored the electronic structure of the 1D-chains we shifted our focus to characterizing isolated domains of 2D-[4]TCOF honeycomb lattices. Figure 3F shows representative d$I$/d$V$ point spectra recorded at the positions marked in the BRSTM inset, the center of a crystalline island comprised of ~10 unit cells (~20 fully cyclodehydrogenated, exclusively corner-linked [4]T units). As previously observed for 1D-chains the dominant spectral features, a broad peak at positive bias ($V_s$ = +1.05 ± 0.05 V, *feature* 4) mirrored by a sharper peak ($V_s$ = –0.47 ± 0.05 V, *feature* 5) and a shoulder ($V_s$ = –0.63 ± 0.05 V, *feature* 5) at negative bias are most prominent if the STM tip is placed along the zigzag edges or at the junction linking two [4]triangulene units. F*eatures* 4 and 5 bracket a semiconducting gap, $E_{g,2D}$ ~ 1.52 V (based on peak-to-peak separation), marginally smaller than the 1D-chain. Differential conductance maps recorded at biases corresponding to *features* 4 and 5 show a familiar nodal pattern; dark featureless [4]triangulene cores lined by two bright lobes at the junction interface and slightly weaker intensity along the zigzag edges (Figures 3G,I, Supporting Information Figure S6). We again used DFT-GGA calculations to connect experimental STS with the predicted electronic structure of a freestanding 2D-[4]TCOF honeycomb lattice. Projections of the LDOS at the energy position of the CB (Figure 3H) and VB (Figure 3J) are in good agreement with d$I$/d$V$ maps attributed to *features* 4 and 5, respectively. The striking correspondence of semiconducting band gaps measured for 1D- and 2D-[4]TCOFs with reported SOMO-SUMO gaps for isolated [4]triangulene cores[24] ($E_{g,0D}$ ~ 1.55 V) along with nearly indistinguishable wavefunction localization patterns irrespective of dimensionality (0D-molecule, 1D-chain, 2D-lattice) suggests only minor contributions from orbital hybridization between [4]triangulene units in the electronic ground state of [4]TCOFs.

**First-principles electronic structure calculation.** Figure 4 shows the electronic band structure and LDOS of a 2D-[4]TCOF honeycomb lattice in the AFM state, calculated using *GW* quasiparticle corrections to the DFT-GGA theory.[38,39,50] The VB and CB complexes bracketing the $E_F$ are formed by sets of flat bands (Figure 4A). The $GW$-corrected quasiparticle band gap for a freestanding 2D-[4]TCOF, $E_{g,2D}^{GW}$ ~ 2.41 eV, is larger (by ~0.9 eV) than the value extracted from STS. This common discrepancy can be attributed to the fact that theory does not account for dielectric screening by the Au(111) substrate, resulting in an overestimation of the experimental band gap.[51] The VB complex splits into a near-degenerate pair of bands (VB and VB−1) separated by a small energy gap (~0.1 eV) from the VB−2. The DOS of the filled states (Figure 4B,C) gives rise to the characteristic signature of a sharp peak (*feature* 5) flanked by a shoulder (*feature* 6) observed in d$I$/d$V$ point spectra (Figure 3F). Total energy calculations within DFT show that an antiferromagnetic (AFM) alignment of spins between the two [4]triangulene cores that form the unit cell ([4]$T_A$ and [4]$T_B$ representing the A and B sublattice sites) of the 2D-[4]TCOF honeycomb lattice is favored as the ground state over both a ferromagnetic (FM) configuration and an alternative spin-unpolarized non-magnetic (NM) configuration (Supporting Information Figure S7, Supporting Information Table S1). The spin polarization energy gain of the AFM configuration is 0.60 eV (0.14 eV) per unit cell compared to the NM (FM) state, indicating large magnetic interaction energies consistent with Lieb's theorem for interacting electrons on a bipartite lattice.[41] The LDOS of up- (blue) and down-spin (red) integrated over the [4]triangulene cores at the A and B sublattice sites, respectively, shows the expected localization of the two spin species in the frontier bands (Figures 4B,C). The VB complex (VB, VB−1, and VB−2) is spin-down polarized at the A-site (Figure 4b) and spin-up polarized at the B-site (Figure 4C). The spin polarization is inverted for the corresponding CB complex (CB, CB+1, and CB+2).

Calculations describing the magnetic ordering between adjacent [4]-triangulene cores in 1D-[4]TCOF chains are summarized in Supporting Information Figure S8A. While the GW predicted band gap of 1D-chains is slightly smaller, $E_{g,1D}^{GW}$ ~ 2.42 eV, the spin polarization energy gain increases to 0.74 eV per unit cell (31 meV per zigzag edge atom), suggesting a strong preference for the AFM ground state in 1D-[4]TCOF chains. The spatial distribution of the local spin density mirrors the picture described for 2D-[4]TCOFs; the spin densities associated with the VB complex (VB, VB−1, and VB−2) on the A-site (Supporting Information Figure S87B) are oppositely oriented than those on the B-site (Supporting Information Figure S8C).

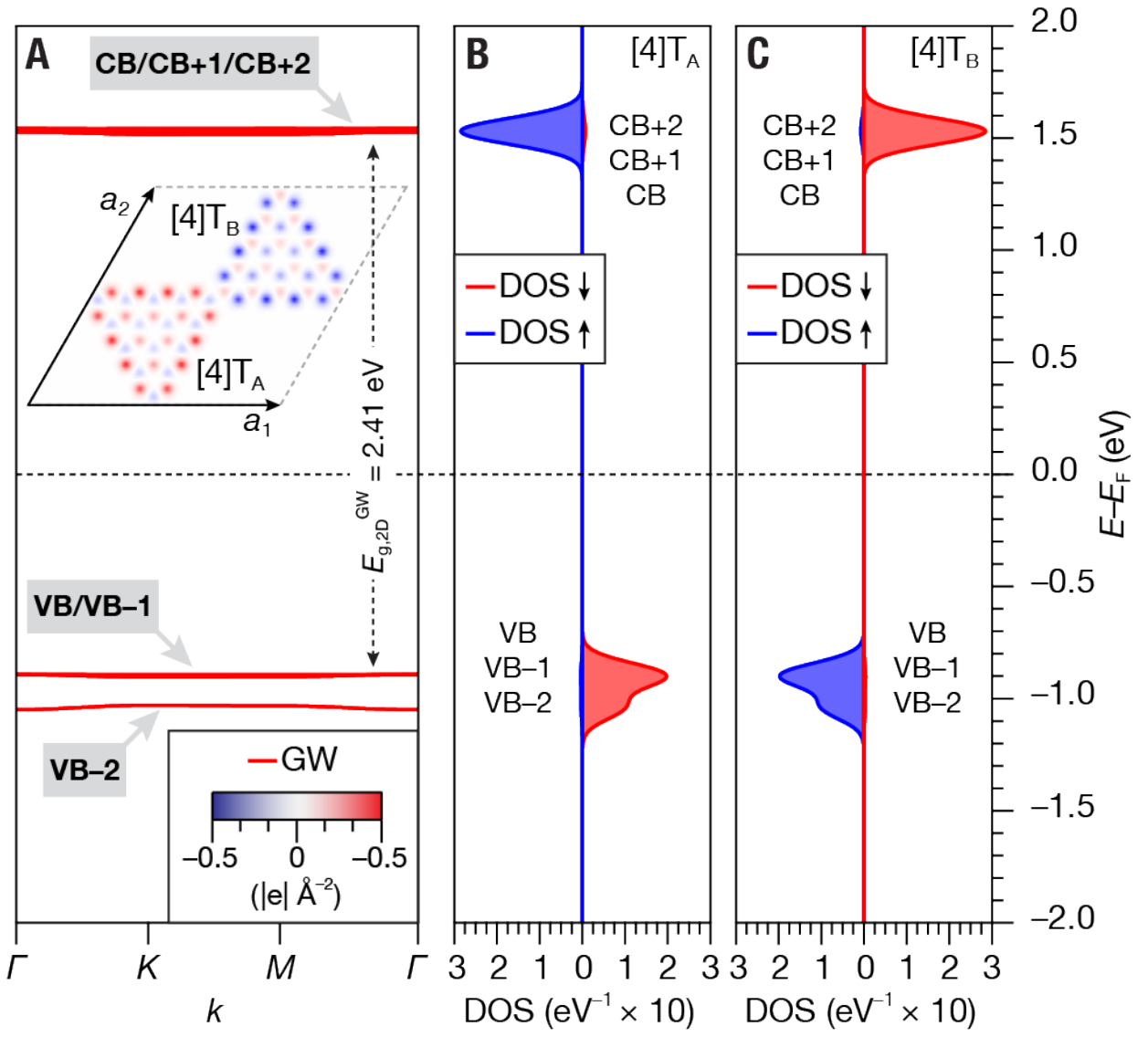

**Figure 4.** Theoretical band structure and sublattice-projected spin-density in 2D-[4]TCOF. (A) GW band structure of a freestanding 2D-[4]TCOF lattice. Inset shows the spatial distribution of the spin density difference ( $\rho(r)^{\uparrow} - \rho(r)^{\downarrow}$) integrated over all occupied states. (B–C) GW LDOS of up (blue) and down (red) spin integrated over the [4]triangulene core at the A ([4]$T_A$) and B ([4]$T_B$) lattice sites, respectively (43 meV Gaussian broadening).

**Low-energy spin excitations in [4]TCOF lattices.** Besides the prominent VB and CB complexes dominating the STS, low bias d$I$/d$V$ spectra (−50 mV < $V_s$ < +50 mV) of 2D-[T]COF lattice and 1D-4[T]COF chains revealed characteristic features commonly associated with spin excitations. Inelastic tunneling spectroscopy (IETS) with metallic STM tips (Supporting Information Figure S9A) recorded at the center or along the zigzag edges of [4]triangulene units embedded in a finite 2D-[4]TCOF honeycomb lattice comprising 24 fully planarized [4]T units (Supporting Information Figure S9C) showed a prominent dip in the conductance signal at zero bias that is absent in isolated [4]triangulene monomers. Either side of $V_s$ = 0 mV, we observed a series of overlapping conductance steps that correspond to inelastic tunneling processes. $d^2I/dV^2$ traces obtained by numerical differentiation[52] revealed the signature of four peaks (corresponding to conductance steps in d$I$/d$V$ spectra) centered at ± 4 mV, ± 13 mV, ± 28 mV, and ± 38 mV (Figure 5A). Their relative intensities vary with the position and bonding environment of the 4[T] core within the honeycomb lattice, trivalent *vs.* divalent *vs.* monovalent substitution, yet remain consistently dominated by the lowest energy mode (Supporting Information Figure S9B).

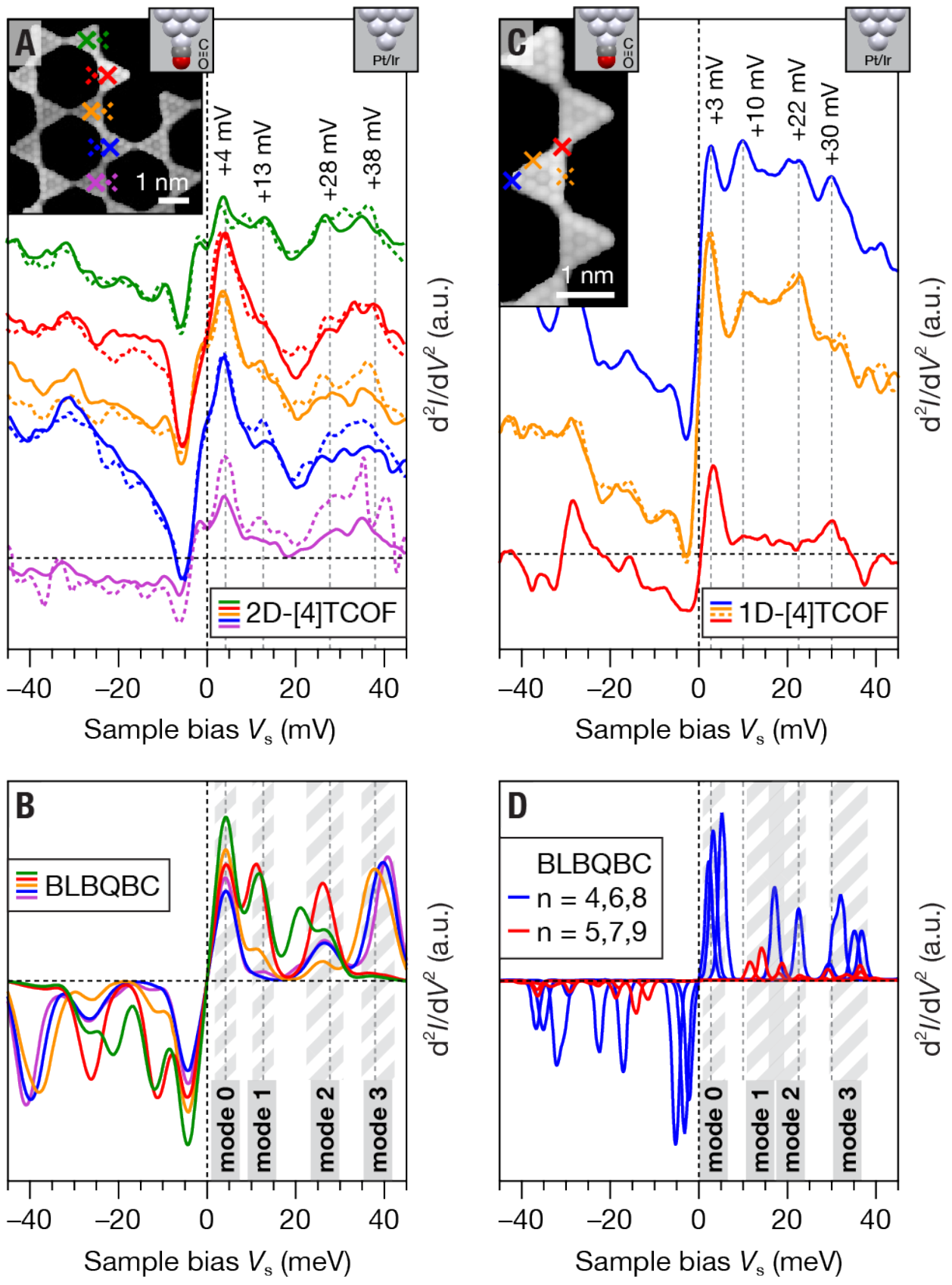

**Figure 5.** Low-energy spin excitations in 2D-[4]TCOF. (A) Low-bias $d^2I/dV^2$ point spectroscopy of a finite 2D-[4]TCOF island comprising 24 [4]T units on Au(111) recorded at the positions marked in the inset (spectroscopy: $V_{ac}$ = 1 mV, $f$ = 455 Hz; metallic STM tip). (B) Spin excitations predicted by a BLBQBC model for an isolated 2D-[4]TCOF

lattice of the same finite size. (C) Low-bias $d^2I/dV^2$ point spectroscopy of a finite-size sample of 1D-[4]TCOF chain of four [4]T units on Au(111) recorded at the positions marked in the inset (spectroscopy: $V_{ac}$ = 1 mV, $f$ = 455 Hz; metallic STM tip). (D) Spin excitations predicted by a BLBQBC model for an isolated 1D-[4]TCOF chains with even (blue lines) and odd (red lines) number of [4]T units. All STM experiments were performed at $T$ = 4.5 K. The theoretical curves are broadened with a 2 meV Gaussian broadening.

To gain further insight into the origin of these low-energy spin excitations, we applied the BLBQBC spin Hamiltonian (eq. 1) to model a finite 2D-[4]TCOF flake (Supporting Information Figure S9C). For the parameters, we take results from earlier theoretical work employing complete active space (CAS) calculations for a 4[T] dimer,[18, 28] and use the biquadratic coefficient $\beta_2$ = 0.06 and bicubic coefficient $\beta_3$ = 0.01, in accordance with CAS Hubbard parameters corresponding to the reported crystalline exchange coefficient $J$ = 12.7 meV from DFT. The simulated $d^2I/dV^2$ traces[53] (Figure 5B) faithfully reproduce the four characteristic features identified in IETS. The lowest energy excitation (*mode* 0) centered at $V_s$ = ± 4 meV corresponds to a nodeless spin excitation mode featuring spectral contributions from all [4]T cores in the finite size 2D-[4]TCOF lattice shown in the insert of Figure 5A. The intensity of *mode* 1 ($V_s$ ~ ± 13 meV) is highly variable and largely associated with isolated [4]T cores fused via one or two corners to a larger 2D-[4]TCOF lattice (mono- or divalent substitution). The energy associated with *modes* 2 ($V_s$ = ± 28 ± 2 meV) and 3 ($V_s$ = ± 38 ± 2 meV) shift slightly with the position in the lattice and have the largest contributions from [4]T cores along the edges (divalent substitution) and the bulk center (trivalent substitution) of the COF lattice, respectively. Notably, the energy of *mode* 3 coincides with the Van Hove singularity predicted for the magnon density of states of the infinite 2D honeycomb lattice,[18] indicating that the excitation spectrum of the finite lattice already captures the magnon physics of the extended system.

While uncontrolled crosslinking of precursor **1b** in dilute networks precludes the characterization of a pristine 1D chain (Supporting Information Figure S4), $d^2I/dV^2$ traces recorded on short segments of 1D-[4]TCOF structures reveal peaks centered at ± 3 mV, ± 10 mV, ± 22 mV, and ± 30 mV (Figure 5C). We herein compared the experimental IETS with theoretical predictions for chain lengths spanning 4–9 units (Figure 5D). The spin excitation spectrum of 1D-[4]TCOFs is sensitive to chain length parity, which determines the ground state's total spin degeneracy (see Methods). Even-length chains feature prominent peaks near $V_s$ ~ ± 3 meV (*mode* 0), $V_s$ ~ ± 20 meV (*mode* 2), and $V_s$ ~ ± 34 meV (*mode* 3) that are all absent for odd lengths, whereas a smaller peak at $V_s$ ~ ± 13 meV (*mode* 1) is only observed in odd-length chains. The dominance of *modes* 0, 2 and 3 in the experimental spectrum points to a chain largely consistent with even parity. The presence of a discernible $V_s$ = ± 10 mV (*mode* 1) feature in some spectra suggests residual odd-parity character, which can be attributed to the influence of defect sites and cross-links that may effectively modulate the active chain length.

## CONCLUSIONS

This work demonstrates that reticular chemistry, combined with on-surface synthesis, enables the deterministic programming of collective magnetic behavior arising solely from π-electron correlations in a carbon-based synthetic lattice. By establishing COFs as a modular and tunable platform, our results open new opportunities to investigate collective quantum phenomena, including lattice-driven magnetic frustration inherent to the honeycomb geometry and the emergence of unconventional magnetic states in organic systems. Extending the synthesis to related frameworks with alternative lattice structures, as well as the incorporation of triangulene homologues with distinct total spin quantum numbers, offers a systematic route to disentangle the interplay between lattice geometry, per-site spin magnitude, and exchange interactions. More broadly, direct access to collective spin excitations and to field- and temperature-dependent magnetic phase transitions in these systems offers a pathway to elucidate quantum magnetism in low-dimensional, molecule-based materials.

## METHODS/EXPERIMENTAL

**Growth of [4]T chains and COFs.** [4]TCOFs were grown on Au(111)/mica films under UHV conditions. Atomically clean Au(111) surfaces were prepared through iterative $Ar^+$ sputter/anneal cycles. 0.9 ML coverage of **1a**,**b** on atomically clean Au(111) was obtained by sublimation at crucible temperatures of 463–473 K using a home-built Knudsen cell evaporator. After deposition the surface temperature was held a 333 K for 12 h before slowly ramping (≤ 2 K $min^{-1}$) to 523 K and held at this temperature for 20 min.

**Scanning Tunnelling Microscopy and Spectroscopy.** All STM experiments were performed using a commercial OMICRON LT-STM operating at $T$ = 4.5 K using PtIr STM tips. $dI/dV$ measurements were recorded with both metallic and CO-functionalized STM tips[54] as indicated, using a lock-in amplifier with a modulation frequency of 455 Hz and a modulation amplitude of $V_{RMS}$ = 10 mV for high-bias experiments and $V_{RMS}$ = 1 mV for low-bias experiments. $dI/dV$ point spectra were recorded under open feedback loop conditions. $dI/dV$ maps were collected under constant current conditions. BRSTM images were obtained by mapping the out-of-phase $dI/dV$ signal collected during a constant-height $dI/dV$ map at 5–10 mV. Peak positions in $dI/dV$ point spectroscopy were determined by fitting the spectra with Lorentzian peaks. Peak positions are calibrated to the Au(111) Shockley surface state. All images were processed using the Gwyddion software package.[55]

**First-principles simulation of 2D-[4]TCOF.** We used Quantum ESPRESSO[50, 56, 57] to determine colinear spin-polarized DFT electronic structure of the [4]TCOF structures under the GGA approximation with a Perdew–Burke–Ernzerhof (PBE) exchange-correlation functional.[42] To this end, we employed a plane wave basis with an 80 Ry cutoff, optimized norm-conserving Vanderbilt pseudopotentials (ONCVPSP),[58, 59] a 6 × 6 × 1 Monkhorst-Pack k-grid,[60] and a 2D Coulomb truncation scheme.[61] Moreover, we used a vacuum spacing of 20 Å in the out-of-plane direction. In this process, we first performed structural and 2D unit cell relaxation until net forces on the atoms were less than $2 \times 10^{-5}$ atomic units and cell pressure less than $10^{-8}$ atomic units. Subsequently, we did a self-consistent charge density computation followed by a final single shot calculation of the DFT electronic structure.

Given that the atomic species are light, we expect the scalar relativistic (SR) formalism to be adequate and that fully relativistic (FR) effects, including spin-orbit coupling (SOC), are negligible. We verified this by comparing the charge density field of SR and FR with SOC calculations and found them to agree to within 0.02%. Furthermore, the computed DFT band structures of the SR and FR with SOC systems differ by at most 1.4 meV for the first six valence and conduction bands each (note that spin is not a good quantum number when comparing to the FR with SOC method).

The subsequent GW calculations were performed with the BerkeleyGW package[38, 39] with 720 unoccupied bands, 6 Ry cutoff for dielectric matrix and screened Coulomb interaction, and 50 Ry cutoff for bare Coulomb interaction.

Inelastic tunneling spectroscopy calculation. For $S = 3/2$, a full solution of the BLBQBC model for a finite system requires a Hilbert space that is $4^N$-dimensional, where $N$ is the number of sites in the system. The finite-size 2D-[4]TCOF of interest here has $N$ = 24 sites, rendering such an approach insoluble. We therefore perform a Holstein-Primakoff transformation, expanding to leading order in $1/S$ to obtain

$$\widehat{\mathcal{H}}_{BLBQBC} \approx J(S - 2\beta_2 S^2 + 3\beta_3 S^3) \sum_{<ij>} \left( a_i^+ a_i + a_j^+ a_j + a_i a_j + a_i^+ a_j^+ \right)$$

Here $a_i, a_j, a_i^+, a_j^+$ are bosonic operators. Following this, we performed symplectic diagonalization for a graph corresponding to the experimental 2D-[4]TCOF domain (Supporting Information Figure S9C) to obtain a bosonic spectrum of the BLBQBC model. We then incorporated the spectrum in an STM-IETS formalism[53] to model the low-bias tunneling current. The first and second derivatives with respect to bias voltage were then determined with finite differences. Here we use the BLBQBC coefficients $J = 12.7$ meV, $\beta_2 = 0.06, \beta_3 = 0.01$ which are consistent with previous theoretical work[14] and lead to results that match well the measured 4, 13, 28, 38 meV peaks.

For first-principles simulation of 1D-[4]TCOF we performed the same DFT+$GW$ procedure as for the 2D-[4]TCOF system, except we configured a cell spacing of 30 Å in the direction orthogonal to the chain but in-plane—in contrast, the 2D-[4]TCOF crystal has a spacing of 19.3 Å in that direction. We also use an 18 × 1 × 1 Monkhorst-Pack $\boldsymbol{k}$-grid. The resultant band structure, spin polarization, and projected density of states (DOS) are summarized in Supporting Information Figure S8. For IETS calculations, we solved the full Hamiltonian instead of the Holstein-Primakoff approximation because the $4^N$-dimensional space is tractable in this case. Moreover, we chose an eigenbasis where total spin and the total $z$-component of spin are good quantum numbers. There the ground state for even chains is $S = 0$ while the ground state for odd chains is $S = 3/2$.

**First-principles simulation of [4]triangulene.** We performed the same DFT+$GW$ procedure as for the 2D-[4]TCOF system, except we used a 25 Å × 25 Å × 15 Å cell and the Martyna-Tuckerman scheme.[62] Furthermore, we sampled only the $\Gamma$ $\boldsymbol{k}$-point. The resultant molecular energies and spin polarization are summarized in Figure 1A,B.

## ASSOCIATED CONTENT

### Supporting Information

The Supporting Information is available free of charge on the ACS Publications website.

Materials and methods, synthetic procedures, STM topographic and BRSTM images, d$I$/d$V$ maps, STS data, NMR spectra

### Code Availability

The DFT code can be downloaded from the Quantum Espresso website [56, 63]. For this study, we used version 7.3 for the DFT-PBE calculations. Elsewhere, the $GW$ code can be obtained from the BerkeleyGW website[48] where we used version 4.0. Finally, we employed ONCVPSP version 0.5 pseudopotentials from the Pseudo Dojo repository[58].

## AUTHOR INFORMATION

### Corresponding Author

* **Felix R. Fischer** - Department of Chemistry, University of California at Berkeley, Berkeley, California 94720, United States; Materials Sciences Division, Lawrence Berkeley National Laboratory, Berkeley, California 94720, United States; Kavli Energy NanoSciences Institute at the University of California, Berkeley and the Lawrence Berkeley National Laboratory, Berkeley, California 94720, United States; Bakar Institute of Digital Materials for the Planet, Division of Computing, Data Science, and Society, University of California Berkeley, Berkeley, California 94720, United States; https://orcid.org/0000-0003-4723-3111; Email: ffischer@berkeley.edu.

* **Steven G. Louie** - Department of Physics, University of California at Berkeley, Berkeley, California 94720, United States; Materials Sciences Division, Lawrence Berkeley National Laboratory, Berkeley, California 94720, United States; https://orcid.org/0000-0003-0622-0170; Email: sglouie@berkeley.edu.

### Authors

**Aidan P. Delgado** - Department of Chemistry, University of California at Berkeley, Berkeley, California 94720, United States; Kavli Energy NanoSciences Institute at the University of California, Berkeley and the Lawrence Berkeley National Laboratory, Berkeley, California 94720, United States; https://orcid.org/0000-0003-0504-6035.

**Carolin Dusold** - Department of Chemistry, University of California at Berkeley, Berkeley, California 94720, United States.

**James Wu** - Department of Physics, University of California at Berkeley, Berkeley, California 94720, United States.

**Jingwei Jiang** - Department of Physics, University of California at Berkeley, Berkeley, California 94720, United States; https://orcid.org/0000-0002-0949-4401.

**Cristina Dadich** - Department of Chemistry, University of California at Berkeley, Berkeley, California 94720, United States; https://orcid.org/0009-0002-3045-2705.

**Adam Cronin** - Department of Chemistry, University of California at Berkeley, Berkeley, California 94720, United States.

### Author Contributions

A.D., C.D., J.W., S.G.L., and F.R.F. initiated and conceived the research. C.D. and C.M.D. designed, synthesized, and characterized the molecular precursors. A.D. and A.C. performed on-surface synthesis and STM characterization. A.D. and F.R.F. analyzed the STM data. J.W., J.J., and S.G.L. developed the theory, performed DFT, GW and spin excitation calculations, and interpreted the data. F.R.F. and S.G.L. acquired the funding and supervised the project. All authors contributed to scientific discussions and writing of the manuscript. A.D., C.D., and J.W. contributed equally to this work.

### Notes

The authors declare no competing financial interest.

## ACKNOWLEDGMENT

This work was primarily supported by: the US Department of Energy (DOE), Office of Science, Basic Energy Sciences (BES), Materials Sciences and Engineering Division under contract DE-AC02-05-CH11231 (F.R.F., S.G.L.) (Nanomachine program KC1203) (molecular design, on-surface synthesis, DFT calculations and tight-binding analyses), the Office of Naval Research (ONR) under contract N00014-24-1-2134 (F.R.F.) (STM imaging and STS characterization) and N00014-19-1-2503 (F.R.F.) (STM instrumentation), and the National Science Foundation under contract DMR-2325410 (S.G.L.) (spin excitations and GW simulations). Support was also provided by the Office of Science, Basic Energy Sciences (BES) under contract DE-SC0023105 (F.R.F.) (molecular precursor synthesis). This research used computational resources provided by the National Energy Research Scientific Computing Center (NERSC), a Department of Energy User Facility using NERSC award No. BES-ERCAP0036104, the Frontera computing project at the Texas Advanced Computing Center (TACC) at The University of Texas at Austin, which is supported by National Science Foundation award No. OAC-1818253, and Stampede3 at TACC through allocation MCA94P030 from the Advanced Cyberinfrastructure Coordination Ecosystem: Services & Support (ACCESS) program, which is supported by U.S. National Science Foundation grants No. 2138259, No. 2138286, No. 2138307, No. 2137603, and No. 2138296 (S.G.L.). A.D. acknowledges support from a Kavli ENSI Graduate Student Fellowships. F.R.F. acknowledges generous support by the Heising-Simons Faculty Fellows Program at UC Berkeley (development of synthetic facilities). We thank Dr. Hasan Çelik and the UC Berkeley NMR facility in the College of Chemistry (CoC-NMR) for assistance with spectroscopic characterization. Instruments in the CoC-NMR are supported in part by National Institutes of Health (NIH) award no. S10OD024998. We thank Dr. Zhongrui Zhou at the UC Berkeley QB3 Chemistry Mass Spectrometry Facility for assistance with mass spectroscopic characterization.

Insert Table of Contents artwork here

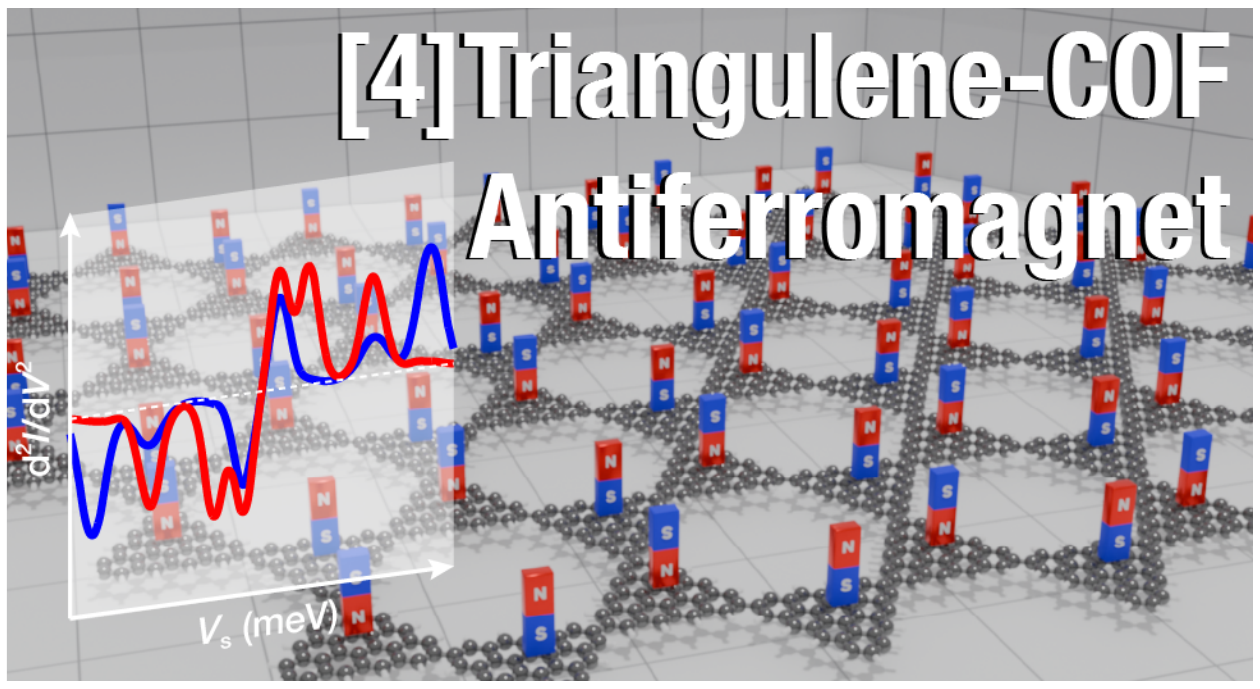